\documentclass[preprint,preprintnumbers,amsmath,amssymb]{revtex4}
\usepackage{graphicx}% Include figure files
\usepackage{dcolumn}% Align table columns on decimal point
\usepackage{bm}% bold math

\begin{document}

\title{Quantum work extraction of a moving battery as a witness to Unruh thermality in high-dimensional spacetimes}
\author{Yan Chen}
\author{Wei-Wei Zhang}
\author{Tian-Xi Ren}
\author{Xiang Hao}
\altaffiliation{Corresponding author}
\email{xhao@mail.usts.edu.cn}
\affiliation{School of Physical Science and Technology, Suzhou University of Science and Technology, Suzhou, Jiangsu 215009, People's Republic of China}
\affiliation{Pacific Institute of Theoretical Physics, Department of Physics and Astronomy,
\\University of British Columbia, 6224 Agriculture Rd., Vancouver B.C., Canada V6T 1Z1.}

\begin{abstract}

We put forward a physical model of a uniformly accelerated Unruh-DeWitt battery and use quantum work extraction as a probe to witness the thermal nature of the Unruh effect in a high dimensional Minkowski spacetime. By means of the open quantum system approach, we investigate the maximal amount of quantum work extraction with respect to the acceleration-induced Unruh temperature, spacetime dimensionality and field mass. It has been found that the steady amount of quantum work extraction in the asymptotic condition is just determined by the Unruh temperature in arbitrary dimensional spacetimes. The asymptotic behavior can demonstrate the global feature of Unruh thermality dependent on the Kubo-Martin-Schwinger condition. From a local viewpoint of Unruh effect, we study the different ways for the dynamics of quantum work extraction when the battery gradually arrives at the same steady state. In the massless scalar field, the evolution with a small acceleration takes on a unique monotonicity in $D=3$ dimensional spacetime and changes to a decaying oscillation for other higher dimensions. The increase in spacetime dimensionality can increase the energy storage capacity of the moving battery. If the mass of the scalar field is considered, the related quantum work extraction is so robust against the Unruh decoherence that the high values can keep for a very long time. The persistence of quantum work extraction is strengthened in higher dimensional spacetime.

\end{abstract}

\maketitle

\section{Introduction}

When paying attention to black hole evaporation, Unruh found that a uniformly accelerating observer in Minkowski spacetime may perceive the ordinary vacuum as a thermal bath of particles with a certain temperature, which is referred to as the Unruh thermal effect \cite{Unruh1976}. This conclusion can be drawn through various approaches and extended to many situations, such as Hawking radiation of black holes \cite{Hawking1975} or particle excitation in inflationary universe \cite{Gibbons1977,XLiu2016}. Aside from the explanation relating Minkowski and Rindler quantizations \cite{Fulling1973,Davies1975}, the Unruh effect can be studied by the response spectrum of a moving Unruh-DeWitt(UDW) detector which consists of a pointlike two-level system coupled to the vacuum field along an accelerated trajectory \cite{Crispino2008}. In particular, the detector in equilibrium with a massless scalar field in the $4$-dimension flat spacetime can exhibit the power spectrum in a Planckian form.

However, the precise connection between the Unruh effect and thermal radiation, as well as its underlying mechanisms, remains an open question in the field. According to the thermalization theorem, an accelerating detector is causally disconnected from the degrees of freedom screened by the global Rindler horizon and presents the loss of information guaranteed by the Kubo-Martin-Schwinger(KMS) condition \cite{Takagi1986}. The thermalization theorem ensures the KMS condition but does not directly imply a Planckian spectrum. For example, in a massive scalar background, the detector's response may differ significantly from the Planck distribution. Recently, a combination of relativity and quantum information theory has opened a new way for capturing the global feature of Unruh thermality. Some measures of quantum resources \cite{Martin2015,Doukas2013,Feng2013,Han2018} and quantum parameter estimation \cite{Hao2019,Chowdhury2022} have been exploited to reveal the thermal nature of the Unruh effect. Nevertheless, a puzzle appears when the number of the spacetime dimensions different from four is taken into account. If the spacetime dimension is even, an accelerating observer would perceive a Bose-Einstein distribution for Bosonic fields and a Fermi-Dirac distribution for Fermionic fields, which are consistent with our intuition \cite{Takagi1985,Louko2016}. On the contrary, Takagi demonstrated the emergence of statistic inversion in the odd dimensional spacetime where the observer could feel a Bose-Einstein distribution for Fermionic fields and a Fermi-Dirac distribution for Bosonic fields. The apparent interchange between Bose-Einstein and Fermi-Dirac distributions was ascribed to the absence of Huygens principle in odd dimensional spacetime \cite{Ooguri1986}. The analytical structure of Wightman function leads to anticommutativity for timelike separated scalar correlator. Some works have emphasized that the spurious statistic inversion can be regarded as a local feature of the accelerating detector \cite{Sriramkumar2003,Arrechea2021}. Although the anomalous statistic inversion could not destroy the thermal character of Unruh effect, which is only dependent on the KMS condition, it is of great value to acquire some physical intuition from the local response.Recent studies have employed geometric phase\cite{Yu2012} and quantum Fisher information\cite{Zhang2022} as alternative probes for the Unruh effect. These tools allow for a deeper exploration of quantum correlations and statistical inversion, offering insights into the thermal and local features of the Unruh radiation that are not immediately visible through conventional thermal characterizations.

In this paper, we suggest a relativistic UDW battery and use quantum work extraction to investigate Unruh thermality from both local and global perspectives\cite{Bruschi2020,Mukherjee2024}. The model of quantum battery has been viewed as quantum system which can exploit quantum resources to store energy from an external field or quantum chargers \cite{Campaioli2017,Le2018,Ferraro2018,Zhang2019,GarciaPintos2020,Farina2019,Ghosh2021,Cruz2022,Barra2022,Rossini2020}. If a quantum battery is in a passive state, no work can be extracted through cyclic unitary operations. The maximal amount of extractable work is defined as the ergotropy which is considered as an elementary quantity for evaluating the charging performance of quantum battery.\cite{Alicki2013} In particular, we explore the ergotropy for the UDW battery, which is driven by an external field and coupled to scalar fields in arbitrary dimensional Minkowski spacetimes. By means of the open quantum system approach, we can obtain the dynamics of quantum work extraction in terms of response functions. In the asymptotic condition, the steady-state values of quantum work extraction can reveal the global side of Unruh thermality. It is interesting to explore the connection of quantum work extraction with the thermal nature of the Unruh effect from the viewpoint of relativistic quantum thermodynamics.

The motivation for exploring Unruh thermality by using quantum work extraction lies in the following factors. Firstly, as an operational estimation of energy transfer between the battery and external environment, quantum work extraction may encode the special trait of the response function which is closely related to the statistical phenomena in the case of different spacetime dimensionality. In some sense, quantum work extraction helps for observing the local side of Unruh effect. Besides it, the behavior of quantum work extraction is also determined by quantum resources of the relativistic quantum battery. It has been proved that nonlocal feature such as quantum coherence can play a role in the performance of the relativistic quantum battery \cite{XHao2023}. If we obtain the dynamics of quantum work extraction, we can demonstrate the evidence of quantum nature of the Unruh effect embodying quantum correlations across Rindler horizon. The nonlocal nature of the Unruh effect is directly connected with the universal thermalization. The attenuation of quantum resources in relativistic thermodynamics results from vacuum fluctuations in spacetimes\cite{Ahmadi2014,Zhao2020,Du2021,Liu2021}.

Our proposal is to construct an accelerated quantum battery that is coupled to a fluctuating vacuum scalar field in an arbitrary-dimensional flat spacetime. We treat the battery as an open quantum system, with the vacuum fluctuations of the scalar field acting as the environment. The evolution of the quantum battery is influenced by decoherence, which arises from the interaction with the massless or massive scalar field during the charging period. From viewpoint of energy transfer, we attempt to probe the thermal properties of the Unruh effect in high-dimensional Minkowski spacetime. The dynamics of quantum work extraction allows us to probe quantum properties of vacuum states, thus providing reliable evidences for the thermal nature of the Unruh effect.

We employ natural units $c=\hbar=1$ throughout the paper. The paper is organized as follows. In Sec. II, we propose a scheme of a Unruh-DeWitt battery and introduce the ergotropy, from the perspective of relativistic quantum thermodynamics. We explore the response function which determines the quantum work extraction. In Sec. III, we study the dynamics of the quantum work extraction, which can consist of the asymptotic behavior and the time-dependent phenomena. The effects of the spacetime dimensionality and field mass on the ergotropy are studied in detail. Finally, in Sec. IV, we give our conclusions and discussions.

\section{Dynamical evolution of UDW battery in a $D$-dimensional Minkowski spacetime}

We put forward a proposal for a UDW battery in a $D$-dimensional Minkowski spacetime.The model of UDW battery is considered as a two-level atom moving along a uniformly accelerated trajectory. The static battery is generated by the Hamiltonian of $H_0=\omega_0 \sigma^{+}\sigma^{-}$ where $\omega_0$ denotes the transition frequency between an excited state $|e\rangle$ and a ground state $|g\rangle$. $\sigma^{\pm}$ represent the rising and lowing operator respectively and the commutator relations of $\sigma^{\pm}$ is $[\sigma^+,\sigma^-]=\sigma_z$. During a charging period, a classical coherent field is applied to drive the battery by the dipolar interaction between the atom and external field in the resonant condition. In the interaction picture, the Hamiltonian of the driven battery is written as $H^{(b)}=\mu(t)\frac {\Omega}2(\sigma^{+}+\sigma^{-})$ where the switching function $\mu(t)=1(0\leq t\leq \tau)$ describes the charging process and $\Omega$ is the effective coupling strength. At $\tau=0$, the battery is prepared in the ground state, which describes the state of the depleted battery. Under the circumstance of no movement, the evolved state of the battery can be governed by $\rho(\tau)=U(\tau)\rho_0 U^{\dag}(\tau)$ where the cyclical unitary operator is $U(\tau)=\mathcal{T}\exp [-i\int_{0}^{\tau} \mathrm{d}s H^{(b)}(s) ]$. Here, the symbol $\mathcal{T}$ denotes the time ordering operator and $\rho_0$ is an initial state. According to \cite{Alicki2013}, the optimal work taken over all unitary transformations $\{ U(\tau)  \}$, i.e., the ergotropy, can be defined as
\begin{equation}
\label{eq:(1)}
\mathcal{W}(\tau)=\mathrm{Tr}[H_0\rho(\tau)]-\min_{\{ U \}}\mathrm{Tr}[U\rho(\tau)U^{\dag}H_0].
\end{equation}
The minimal unitary transformation $U_{\sigma}$ satisfies that the states $U_{\sigma}\rho(\tau)U^{\dag}_{\sigma}=\sum_{j}\varrho_j|\epsilon_j\rangle  \langle \epsilon_j | $ are passive. Here, $|\epsilon_j\rangle $ is the energy level state of $H_0$ with the corresponding energy $\epsilon_j$ in the increasing order, $\epsilon_j<\epsilon_{j+1}$ and $\varrho_j$ is the eigenvalues of $\rho(\tau)=\sum_j \varrho_j |\varrho_j\rangle \langle \varrho_j|$ in the decreasing order. Therefore, the optimal unitary operation $U_{\sigma}=\sum_j |\epsilon_j\rangle \langle \varrho_j|$ is exploited to obtain the ergotropy in the form of
\begin{equation}
\label{eq:(2)}
\mathcal{W}(\tau)=\sum_{j,k}\varrho_j \epsilon_k (|\langle \varrho_j|\epsilon_k\rangle|^2-\delta_{jk}).
\end{equation}
According to the second law of thermodynamics, the energy stored in the battery cannot be wholly extracted by the cyclic unitary transformation. The larger maximal amount of extractable work, the better charging performance of quantum battery. In fact, quantum battery is inevitably influenced by the surrounding environment. The ergotropy of quantum battery can also be constrained by quantum decoherence from the environment.

Considering the uniformly accelerated motion, we treat the battery as an open quantum system which is coupled to a bath of fluctuating quantum scalar field in a $D$-dimensional Minkowski spacetime. Its density matrix is governed by the Lindblad form of the master equation and undergoes quantum decoherence and dissipation. The total Hamiltonian of the combined system of quantum battery and scalar fields in a $D$-dimensional flat spacetime can be written as
\begin{equation}
\label{eq:(3)}
H=H^{(b)} + H^{(\phi )} + H^{(I)}.
\end{equation}
The Hamiltonian for the free scalar field ${H^{(\phi )}}$ is defined for the scalar field $\Phi \left( x \right)$, which satisfies the standard Klein-Gordon equation in $D$-dimensional Minkowski spacetime. ${H^{(I)}} =\mu \lambda ({\sigma ^ + } + {\sigma ^ - })\Phi \left( {x(\tau )} \right)$ represents the interaction between the battery and scalar field, where $\mathbf{\Phi} \big( x(\tau) \big)$ corresponds to the scalar field operator and $\lambda \ll \Omega$ characterizes the weak coupling constant. It is seen that the UDW battery can be charged by both the external driving field and the quantum scalar field.

In the condition of weak couplings, the initial state of combined system can be approximated as $\rho_{tot}(0)=|g\rangle\otimes |0\rangle \langle 0|$, where $\rho(0)$ is the initial state of the atom and $|0\rangle$ denotes the vacuum state of scalar field in a $D$-dimensional Minkowski spacetime. In the frame of the moving battery, the reduced density matrix $\rho(\tau)$ of the battery can be obtained by the quantum master equation in the Kossakowski-Lindblad form of
\begin{align}
\label{eq:(4)}
\frac {\partial}{\partial \tau} \rho(\tau)&\;=\;-i[H^{(b)}_{eff}, \rho(\tau)]+\frac 12\sum_{i,j=1}^{3}a_{ij}\mathcal{D}_{ij}[\rho(\tau)], \\
a_{ij}&\;=\;A\delta_{ij}-iB\varepsilon_{ijk}\delta_{k1}+C\delta_{i1}\delta_{j1}, \nonumber \\
A&\;=\; \frac {\lambda^2}{2}[\mathcal{G}(\Omega)+\mathcal{G}(-\Omega)],\;B= \frac {\lambda^2}{2}[\mathcal{G}(\Omega)-\mathcal{G}(-\Omega)],\;C=\lambda^2 \mathcal{G}(0)-A,\nonumber
\end{align}
where the dissipator $\mathcal{D}_{ij}(\rho)=2\sigma_j\rho\sigma_i-\sigma_i\sigma_j\rho-\rho\sigma_i\sigma_j$ arises from the dissipation and decoherence induced by the environment. $\{ \sigma_j,(j=1,2,3)\}$ are the three components of Pauli operators. The Kossakowski matrix $a_{ij}$ can be explicitly resolved. By introducing the Wightman function of scalar field $ G^{+}(x-x')=\langle 0|\Phi \big( x(\tau) \big)\Phi \big( x'(\tau^{'}) \big)|0\rangle=\frac {1}{4\pi^2[|\vec{x}-\vec{x}'|^2-(t-t'-i\epsilon)]}$, we can derive its Fourier transform
\begin{equation}
\label{eq:(5)}
\mathcal{G}(\Omega)=\int_{-\infty}^{\infty} \mathrm{d}\Delta \tau \cdot e^{i\Omega\Delta \tau} G^{+}(\Delta \tau).
\end{equation}
The Hilbert transform of the Wightman function is given by $\mathcal{K}(\Omega)=\frac {\mathcal{P}}{\pi i}\int_{-\infty}^{\infty} \mathrm{d}\omega\frac {\mathcal{G}(\omega)}{\omega-\Omega}$
where $\Delta \tau=\tau-\tau'$ and $\mathcal{P}$ denotes the principle value. The effective Hamiltonian is given by $H^{(b)}_{eff}=\frac {1}{2}\Omega^{'}(\sigma^{+}+\sigma^{-})$ with $\Omega^{'}=\Omega+i\lambda^2[\mathcal{K}(-\Omega)-\mathcal{K}(\Omega)]$ representing the effective coupling. The interaction with external scalar field would have an effect on the Lamb shift. In the case of weak couplings, $\lambda^2 \ll \Omega$, we can neglect the Lamb shift in the following.

With respect to a uniformly accelerated battery in a $D$-dimensional Minkowski spacetime, it is known that the field Wightman function fulfills the KMS condition i.e., $G^+(\Delta\tau) = G^+(\Delta\tau+ i\beta)$. Equivalently, in the frequency space, the KMS condition can be demonstrated by $\mathcal{G}(\lambda) = e^{\beta \Omega}\mathcal{G}(-\lambda)$ where $\beta=1/T_U$ represents the Unruh temperatute. To proceed, we should explore the dynamics of the UDW battery. For a two-level atom, the density matrix  $\rho(\tau)$ can be expressed in the form of
$\rho(\tau)=\frac {I+\sum_j r_j(\tau)\sigma_j}{2}$ where $r_j=\mathrm{Tr}(\sigma_j \rho)$ is the $j$th-component of the Bloch vector. Therefore, the dynamics of the UDW battery will satisfy the Bloch equation,
\begin{equation}
\label{eq:(6)}
\frac {\mathrm{d}}{\mathrm{d}\tau}\mathbf{r}(\tau)=-2\mathcal{H}\cdot \mathbf{r}(\tau)+\mathbf{\chi},
\end{equation}
where the decaying matrix
\begin{equation}
\mathcal{H}=\left(\begin{array}{ccc} 2A & 0 & 0 \\ 0 & 2A+C & \Omega/2 \\ 0 & -\Omega/2 & 2A+C  \end{array}  \right). \nonumber
\end{equation}
and $\mathbf{\chi}=(-4B, 0, 0)^{\mathrm{T}}$ is the inhomogenous vector. We use a quantum channel to describe the dynamics of the battery by mapping the Bloch vector,
\begin{equation}
\label{eq:(7)}
\mathbf{r}(\tau)=\Theta(\tau)\cdot \mathbf{r}(0)+\mathbf{\Lambda}(\tau),
\end{equation}
where \begin{equation}
\Theta(\tau) = \exp(-2\mathcal{H}\tau)=\left(\begin{array}{ccc} e^{-4A\tau} & 0 & 0 \\ 0 & e^{-2(2A+C)\tau}\cos \Omega \tau & -e^{-2(2A+C)\tau}\sin \Omega \tau \\ 0 & e^{-2(2A+C)\tau}\sin \Omega \tau & e^{-2(2A+C)\tau}\cos \Omega \tau  \end{array}  \right) \nonumber
\end{equation}
denotes the mapping matrix of the quantum channel and $\mathbf{\Lambda}(\tau) = \frac 12[I-\Theta(\tau)]\mathcal{H}^{-1}\cdot \mathbf{\chi}$ is the mapping vector.

For a general initial state $\left| \psi  \right\rangle  = \sin \frac{\theta }{2}\left| g \right\rangle  + \cos \frac{\theta }{2}\left| e \right\rangle $, the expression of the Bloch vector for the UDW battery is obtained,
\begin{equation}
\label{eq:(8)}
\mathbf{r}(\tau ) = \left( {\begin{array}{*{20}{c}}
		{\left( {\gamma + \sin \theta } \right){e^{ - 4A\tau }} - \gamma}\\
		{ - {e^{ - 2(2A + C)\tau }}\sin \Omega \tau \cos \theta }\\
		{{e^{ - 2(2A + C)\tau }}\cos \Omega \tau \cos \theta }
\end{array}} \right),
\end{equation}
where the ratio $\gamma=\frac {B}{A}$ is determined by the Unruh temperature due to the frequency KMS condition. It is found that in the asymptotic limit of $\tau \rightarrow \infty$, the state of the battery arrives at the steady state of $\mathbf{r}_s=-\gamma(1,0,0)^{\mathrm{T}}$ which is related to Unruh temperature.

To derive the complete dynamics of the UDW battery, we need to specify $\mathcal{G}(\Omega)$ following the trajectory of the atom with a constant acceleration $a$ in a $D$-dimensional flat spacetime
\begin{equation}
\label{eq:(9)}
{x^0}\left( \tau  \right) = {a^{ - 1}}\sinh a\tau ,{x^1}\left( \tau  \right) = {a^{ - 1}}\cosh a\tau ,{x^2}\left( \tau  \right) = {x^3}\left( \tau  \right) =  \cdots  = {x^{D - 1}}\left( \tau  \right) = 0.
\end{equation}
The quantum scalar field can be expanded as
\begin{equation}
\label{eq:(10)}
\mathbf{\Phi} \left( x \right) = \int {{d^{D - 1}}k\left\{ {{a_k}{U_k}(x) + a_k^\dag U_k^*(x)} \right\}},
\end{equation}
where ${U_k} = {\left[ {2{\Omega _k}{{\left( {2\pi } \right)}^{D - 1}}} \right]^{ - 1/2}}\exp \left[ {i\left( {{k_1}{x^1} + {\bf{k}}\cdot{\bf{x}} - {\Omega _k}{x^0}} \right)} \right]$ is the positive field mode. $\Omega _k^2 = {m^2} + k_1^2 + {\left| {\bf{k}} \right|^2}$ and the field operators obey canonical communication relation $\left[ {{a_k},a_{k'}^\dag } \right] = {\delta ^{D - 1}}(k - k') $. The symbol $m$ denotes the mass of the scalar field.

 Substituting Eqs.(9) and (10) into $\mathcal{G}$, for the UDW battery, we can formally obtain
\begin{equation}
\label{eq:(11)}
{{\mathcal{G_D}^{(\text{m})}(\Omega)}} =  - \frac{\pi }{\Omega}\frac{{{\mathcal{F}_D}(\Omega)}}{{{e^{ - \beta \Omega}} - 1}}.
\end{equation}
Here the profile function is given by ${\mathcal{F}_D}(\Omega) = \frac{2}{{\pi |\Gamma (i\Omega/a){|^2}}}\int {\frac{{{d^{D - 2}}k}}{{{{(2\pi )}^{D - 2}}}}|{K_{i\Omega/a}}(\sqrt {{m^2} + |{\bf{k}}{|^2}/a} )} {|^2}$ where $K_{i\Omega/a}$ is the modified Bessel function .

For a free massless scalar field in D-dimension, Eq. (11) has an analytical form
\begin{equation}
\label{eq:(12)}
{{\mathcal{G_D}^{(0)}(\Omega)}} = \frac{{{\pi ^{\frac{D}{2} - 2}}{\beta ^{3 - D}}\Gamma (\frac{D}{2} - 1)}}{{4\Gamma (D - 2)}}\frac{{{f_D}(\Omega)}}{{{e^{\beta \Omega}} - {{( - 1)}^D}}}
\end{equation}
where
 \begin{equation}
{f_D}(\Omega) = \left\{ \begin{array}{l}
	\frac{{2\pi }}{{\beta \Omega}}\prod\limits_{l = 0}^{(D - 4)/2} {[{l^2} + {{(\frac{{\beta \Omega}}{{2\pi }})}^2}]{\rm{\quad  \quad \ \; \text{if D is even,}}}} \\
	\prod\limits_{l = 0}^{(D - 5)/2} {[{{(l + \frac{1}{2})}^2} + {{(\frac{{\beta \Omega}}{{2\pi }})}^2}]{\rm{\quad\  \text{if D is odd.}}}}
\end{array} \right. \nonumber
\end{equation}
and $\Gamma(\alpha)$ is the Gamma function. For even $D$, a Plankian factor with Bose-Einstein distribution is observed. The Fermi-Dirac distribution is also demonstrated in the case of the odd dimension.

For a free massive scalar field, the integral $\mathcal{F}_D(\Omega) $ admits no simple analytic expression. In particular, for a large field mass with $m \gg {T_U}$ , using the asymptotic form of the modified Bessel function for large argument, the profile function $\mathcal{F}_D(\Omega)  $ can be given by
\begin{equation}
\label{eq:(13)}
{{\mathcal{G_D}^{(\text{massive})}(\Omega)}}\approx \frac{{{m^{D/2 - 2}}{e^{ - m\beta /\pi }}}}{{{2^{D/2 - 1}}{\beta ^{D/2 - 1}}}}{e^{\beta \Omega/2}}.
\end{equation}
Although the above equation does not contain Planck factor, it is still thermal in the sense that it keeps the frequency KMS condition, i.e.,$\mathcal{G}_D^{(\text{massive})}( - \Omega) = {e^{ - \beta \Omega}}\mathcal{G}_D^{(\text{massive})}(\Omega) $.

We can obtain the Kossakowski coefficients for the scalar field with a large mass
\begin{equation}
\label{eq:(14)}
A_D^{(\text{massive})} = \frac{{{m^{D/2}}{e^{ - m\beta /\pi }}}}{{{2^{D/2}}{\beta ^{D/2 - 1}}}}\cosh (\beta \Omega /2),
\end{equation}
and for a massless scalar field
\begin{equation}
\label{eq:(15)}
A_D^{(0)} = \frac{{{\pi ^{{\textstyle{{D - 5} \over 2}}}}{\beta ^{3 - D}}}}{{4\Gamma ({\textstyle{{D - 1} \over 2}})}}{\left| {\Gamma \left( {\frac{D}{2} - 1 + \frac{{\beta \Omega }}{{2\pi }}i} \right)} \right|^2}\cosh (\beta \Omega /2).
\end{equation}
Using Euler reflection formula ${\left| {\Gamma \left( {1/2 + ix} \right)} \right|^2} = \pi /\cosh (\pi x)$ and recurrence relation $\Gamma(z+1)=z\Gamma(z) $, we calculate the decaying coefficients in some simple cases of
\begin{equation}
\label{eq:(16)}
\begin{array}{cc} A_3^{m = 0} = \frac{1}{4}, & A_4^{m = 0} = \frac{{{\Omega }}}{{4\pi }}{\gamma^{ - 1}}, \\ A_5^{m = 0} = \frac{\pi }{{16{\beta ^2}}} + \frac{{\Omega ^2}}{{16\pi }}, & A_6^{m = 0} = \left( {\frac{{{\Omega }}}{{6{\beta ^2}}} + \frac{{\Omega ^3}}{{24{\pi ^2}}}} \right){\gamma^{ - 1}}.  \end{array}
\end{equation}
The hyperbolic factor $\gamma={\tanh(\beta \Omega/2)}$ appears only for even D, which is inherited from the statistics inversion. We observe that the result for $D=3$ is unique, as its Kossakowski coefficient remains constant, whereas in other dimensions, the coefficient is modulated by a polynomial on $\beta$. As we will later demonstrate, this distinct character of $D = 3$ leads to a significantly different evolution of quantum work extraction compared to models in other higher dimensions.

\section{The effects of spacetime dimensionality and field mass on quantum work extraction}

This study aims to explore quantum work extraction of the UDW battery described in Eq. (8), which is used to witness Unruh thermality in a $D$-dimensional Minkowski spacetime. Since the steady state depends only on the Unruh temperature, we expect the asymptotic behavior of quantum work extraction to inherently reveal the global thermal nature of the Unruh effect, in accordance with the KMS condition. We propose that the various evolutions of the ergotropy can highlight the local features of the Unruh effect, which stem from the responses of the UDW detector interacting with a scalar background. Furthermore, we will examine how to effectively amplify the ergotropy by properly varying the spacetime dimensionality and field mass. In this section, we assume that the initial state of the UDW battery is prepared at the ground state.

To obtain the maximal amount of quantum work extraction, we should firstly calculate the eigenvalues $\varrho_{1,2}=\textstyle{{1 \pm \left| {\mathbf{r}(\tau )} \right|} \over 2}$ of the density matrix of the battery. The optimal unitary operation is expressed as $U_\sigma =\sum\nolimits_{j=1,2} {\left|  \epsilon_j\right\rangle \left\langle \varrho_j \right|}$ where $\left|  \varrho_j\right\rangle$ is the eigenvector of the density matrix of the battery state. In the space of $\{|\epsilon_1\rangle=|g\rangle ,|\epsilon_2\rangle=|e\rangle\}$, $|\varrho_{1}\rangle=\sqrt{\frac {r+r_3}{2r}}|g\rangle+\frac {r_1+ir_2}{\sqrt{2r(r+r_3)}}|e\rangle$ and $|\varrho_{2}\rangle=\frac {r_1-ir_2}{\sqrt{2r(r+r_3)}}|g\rangle-\sqrt{\frac {r+r_3}{2r}}|e\rangle$ are obtained. The maximal work extraction can be achieved by the optimal unitary transformation,
\begin{equation}
\label{eq:(17)}
U_{\sigma}=\left( \begin{array}{cc} \sqrt{\dfrac {r+r_3}{2r}} & \dfrac {r_1-ir_2}{\sqrt{2r(r+r_3)}} \\ \dfrac {r_1+ir_2}{\sqrt{2r(r+r_3)}} & - \sqrt{\dfrac {r+r_3}{2r}}  \end{array}  \right),
\end{equation}
where $r=\left| {\mathbf{r}(\tau )} \right|$ represents the norm of the Bloch vector. The value of the ergotropy $\mathcal{W}$ is determined by the internal energy $E(\tau)=\mathrm{Tr}[H_0\rho(\tau)]=\frac {\omega_0(1+r_3)}2$ and the part of $\mathrm{Tr}[U_{\sigma}\rho(\tau)U_{\sigma}^{\dag}H_0]=\mathrm{Tr}[\rho_{\sigma}H_0]=\sum_{j=1,2}\varrho_j \epsilon_j= \frac {\omega_0(1-r)}2$. Therefore, the ergotropy is expressed as $\mathcal{W}=\frac {\omega_0(r+r_3)}2$. For convenience, we can define the scaled ergotropy as $\xi(\tau)=\frac {\mathcal{W}}{\omega_0}$,
\begin{equation}
\label{eq:(18)}
\xi(\tau)=\frac {1}{2}[r(\tau)+r_3(\tau)].
\end{equation}
We emphasize that the UDW battery in our scheme is not only charged by the external driving field, but also by the coupling to the scalar field in a $D$-dimensional Minkowski spacetime. With no external driving field, $\Omega=0$, the energy of the battery is just supplied by Unruh thermality and then the asymptotic ergotropy is null because of $r(\infty)=-r_3(\infty)$ determined by the thermal equilibrium state. In the following study, we will work with dimensionless parameters by rescaling the Unruh temperature and field mass as
\begin{equation}
\label{eq:(19)}
\beta  \mapsto \tilde \beta  \equiv \beta \Omega ,m \mapsto \tilde m \equiv m/\Omega
,\tau  \mapsto \tilde \tau  \equiv {\mu _D}\tau
\end{equation}
where $\mu_D=\lambda^2\Omega^{D-3}$. For convenience, we continue to term $\tilde \beta $, $\tilde \tau$ and $\tilde m$ as $\beta$, $\tau$ and $m$, respectively.

After evolving for enough long time, the ergotropy of the steady state in the asymptotic limit is obtained in the form of
\begin{equation}
\label{eq:(20)}
\xi_s=\frac{\gamma}2.
\end{equation}
As depicted in Fig. 1, it is shown that the asymptotic ergotropy is independent of the scalar field background, which means that the asymptotic value of quantum work extraction is only determined by acceleration-induced Unruh temperatures. The ergotropy gradually decreases with the increase of Unruh temperature, which demonstrates that the higher charging performance of the UDW battery can be achieved in the condition of the smaller acceleration. It is reasonable that quantum decoherence at a low Unruh temperature has weak impacts on quantum work extraction. This result indicates that quantum work extraction primarily captures the global thermal nature of the Unruh effect.

On the other hand, we estimate the dynamics of the ergotropy for the UDW battery within a massless field background. In arbitrary $D$-dimensional flat spacetime, the response function $\mathcal{G}_D$ exhibits the inversion of statistics. Our interest is to investigate how the dimensionality of spacetime influences the evolution of the ergotropy. We expect to find out the different specialities for the evolutions of quantum work extraction in various dimensional spacetimes. Without loss of generality, we focus on some specific cases where the dimensions $D = 3, 4, 5, 6$ are chosen. In Figs. 2, we illustrate the time-dependent behavior of the ergotropy as a function of the Unruh temperature in an even or odd dimensional Minkowski spacetime.

The particular behavior of the ergotropy is shown in Fig. 2(a). When the charging time grows, the ergotropy increases and monotonically approaches an asymptotic value for a certain Unruh temperature. The kind of monotonic phenomena arises from the $\beta$-independence of the Kossakowski coefficient, i.e., ${\partial _\beta }{A_3^{(0)}} = 0$. The ergotropy can be enhanced to a maximum during a short charging period. This result helps for designing the battery with a rapidly charging performance. With respect to other higher dimensions of $D = 4, 5, 6$, we observe that the ergotropy experiences the non-monotonically evolution which resembles a damping oscillation. Comparing the ergotropy evolution and the local response, a remarkable difference between them needs to be emphasized, where no inversion behavior of quantum work extraction as a function of the dimensionality was found. It is understood that the dynamics of quantum work extraction essentially distinguishes the way of the battery thermalization, which is independent of statistical inversion encoded in the response function. As previously mentioned, this non-monotonic behavior of the ergotropy arises directly from the polynomial dependence on $\beta$ in the Kossakowski coefficient, and thus holds true for all models involving a massless scalar background in $D>3$ dimensions. The peaks of the ergotropy can decline with time and arrive at a non-zero steady value after a long time. For the higher dimension, the maximal values of the ergotropy are larger. It is demonstrated that much more energy may be extracted from the UDW battery in the high-dimensional flat spacetime. However, the minimal time for steadily charging is enlarged. It is noteworthy that both the location and magnitude of the ergotropy peak can serve as indicators to distinguish the different spacetime dimension. Unlike the pronounced monotonic behavior observed in $D=3$, the ergotropy in higher-dimensional spacetimes is more sensitive to the Unruh temperature and reaches its maximum at an earlier time, which illustrates a sharper peak during the temporal evolution. Numerical analysis further reveals that the peak value of the ergotropy increases as the dimensionality of spacetime grows.

For a massive scalar field background, the Kossakowski coefficients are mass-dependent. This point means that the related ergotropy may encode certain mass effects of the scalar field. Due to a factor of $e^{-m\beta/\pi }$ appearing in the local response, the mass effect is that the Kossakowski coefficients are modulated by a polynomial on Unruh temperature. Correspondingly, the dynamical behavior of the ergotropy undergoes the non-monotonic oscillation in arbitrary dimensional spacetime.

To further investigate the mass effect, we depict the dynamics of the ergotropy for $m/\omega_0=10$ massive background with a fixed Unruh temperature $T_U/\Omega=0.1$ in Fig. 3. In the large field mass limit of $m\ll T_U$, we observe that the non-monotonic ergotropy can reach some peak values which are much larger than the asymptotic value. It is seen that a significant improvement of the charging performance can be established in the mass field background. With the increase of the spacetime dimension, the enhanced effect is more noticeable. Moreover, the mass-dependent ergotropy exhibits the remarkable robustness against environmental decoherence, as its peak persists for a significantly longer time compared to the massless scalar field. This robustness can be attributed to the fact that the local response of the UDW battery is suppressed by an exponential factor of the mass, causing the battery to attain to the equilibrium state during a much longer period \cite{Takagi1986}. To illustrate this, we have numerically calculated the ergotropy for scalar field backgrounds with different dimension in Fig. 3. The cases of $D=3,6$ are represented by solid and dot-dashed lines, respectively. Our analysis reveals that the persistence of quantum work extraction is enhanced in higher-dimensional spacetimes, while its maximum value also increases.

\section{Discussion}
We have put forward the model of a relativistic UDW battery. We have reexamined the thermal nature of the Unruh effect by employing quantum work extraction as an effective witness. We consider a uniformly accelerated atom driven by an external field, modelled as an open quantum battery coupled to the scalar field background in a $D$-dimensional Minkowski spacetime. The dynamics of the UDW battery can be described by the Lindblad master equation dependent on the local response function. The evolution of quantum work extraction is related to the Unruh temperature, as well as the inherent traits of the scalar background, such as field mass and spacetime dimensionality. Our findings show that the asymptotic ergotropy for the steady state, is independent of the local response but entirely determined by the Unruh temperature. The asymptotic phenomenon encapsulates the global thermal nature of the Unruh effect, as dictated by the KMS condition. On the other hand, the time evolution of the ergotropy captures the local side of Unruh thermality, i.e., the different ways to the same thermal equilibrium. For a massless scalar background with a small acceleration, the ergotropy maintains monotonicity in $D=3$ dimensions, while displaying non-monotonic behavior for $D>3$ higher dimensions. Furthermore, if the large mass of the scalar field is considered, the related quantum work extraction is so robust against the Unruh decoherence that the high values can keep for a very long time. The persistence of quantum work extraction is strengthened in the higher dimension spacetime.

\begin{acknowledgments}
We would like to thank Professor Bill Unruh and Professor Philip C. E. Stamp for the discussions on the related work. This work is supported by Postgraduate Research and Practice Innovation Program of Jiangsu Province(Grant No. KYCX24-3424). $\mathrm{SCOAP}^{3}$ supports the goals of the International Year of Basic Sciences for Sustainable Development.
\end{acknowledgments}

\newpage

{\large \bf Figure Captions}

\vskip 0.5cm

{\bf Figure 1.}

The asymptotic behaviour of the ergotropy for the steady state of an accelerated UDW battery, which is independent of its local response to the vacuum field and demonstrates the global Unruh thermal nature in an arbitrary dimensional spacetime.

\vskip 0.5cm

{\bf Figure 2.}

The evolution of the ergotropy for the UDW battery coupled to a massless scalar field in $D=3,4,5,6$-dimensional Minkowski sapcetime, as a function of the scaled proper time $\tau$ and the Unruh temperature $T_{U}/\Omega$.

\vskip 0.5cm

{\bf Figure 3.}

The robust and oscillating dynamics of the ergotropy for the UDW battery interacting with a large massive field in $D=3,6$ dimensional spacetime. The black solid line represents the case of $D=3$-dimension Minkowski spacetime and the blue dashed line denotes that of $D=6$-dimension spacetime.

\newpage
\begin{figure}
  \centering
  \includegraphics[width=1.0\textwidth]{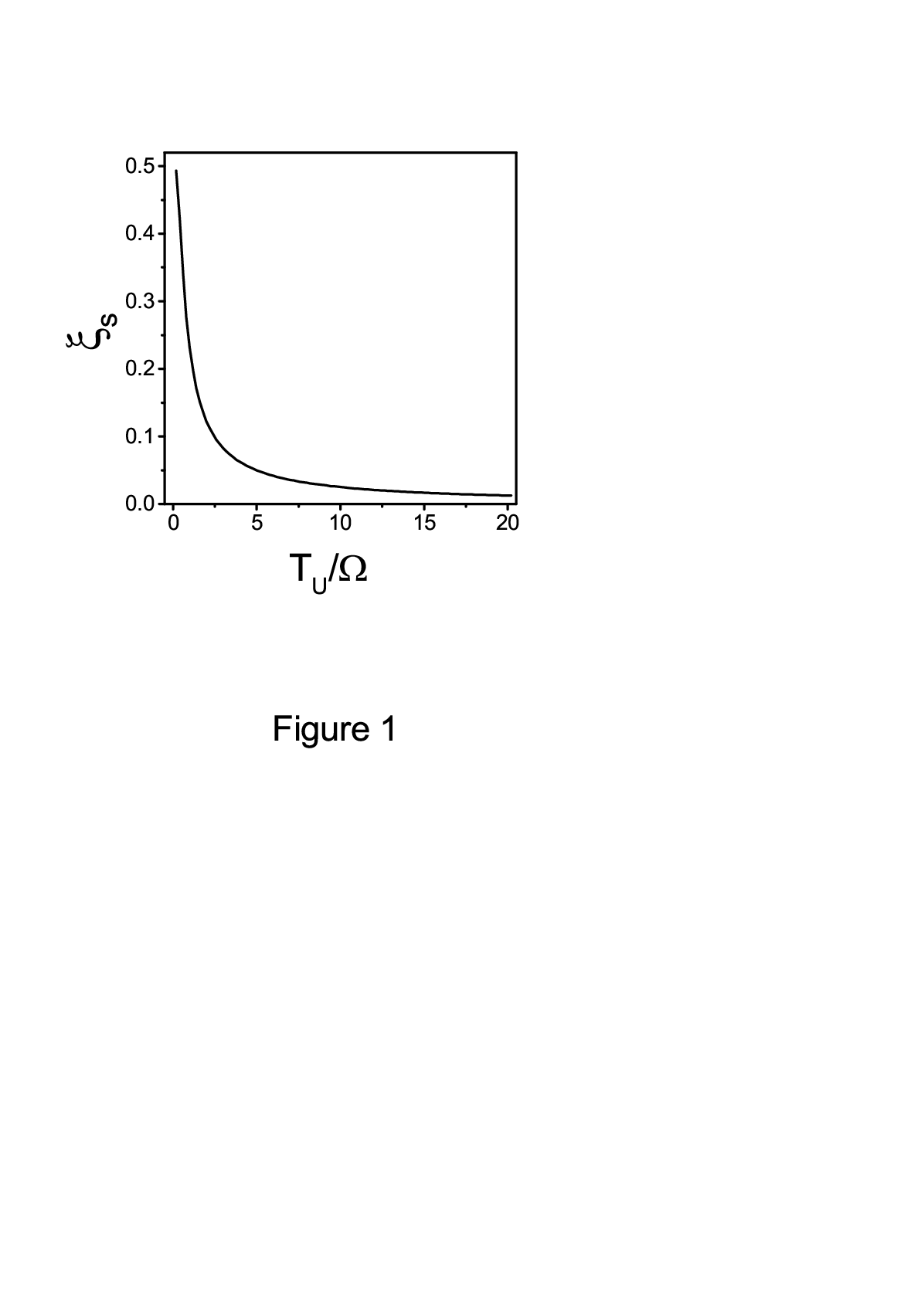}
\end{figure}

\begin{figure}
  \centering
  \includegraphics[width=1.0\textwidth]{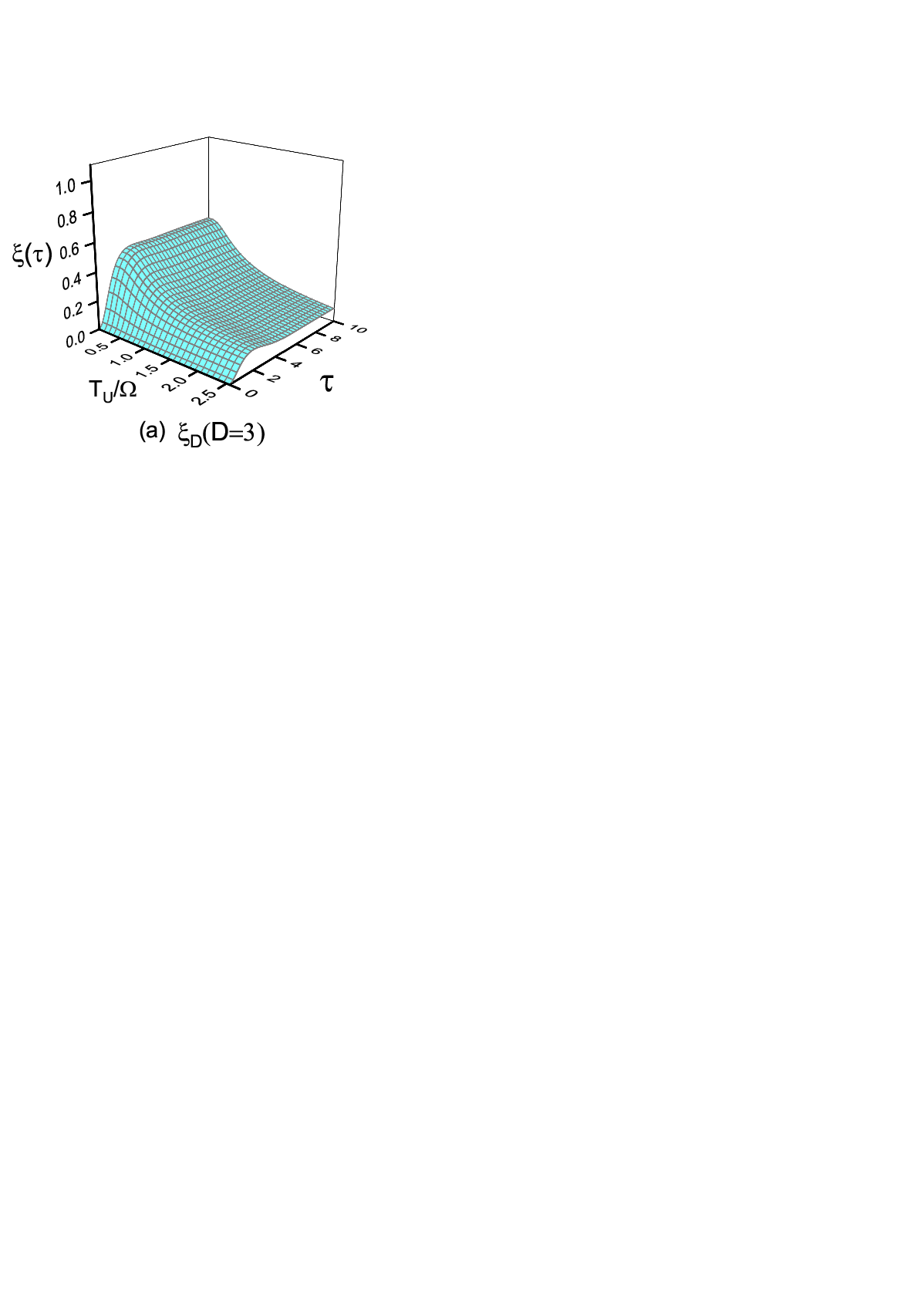}
\end{figure}

\begin{figure}
  \centering
  \includegraphics[width=1.0\textwidth]{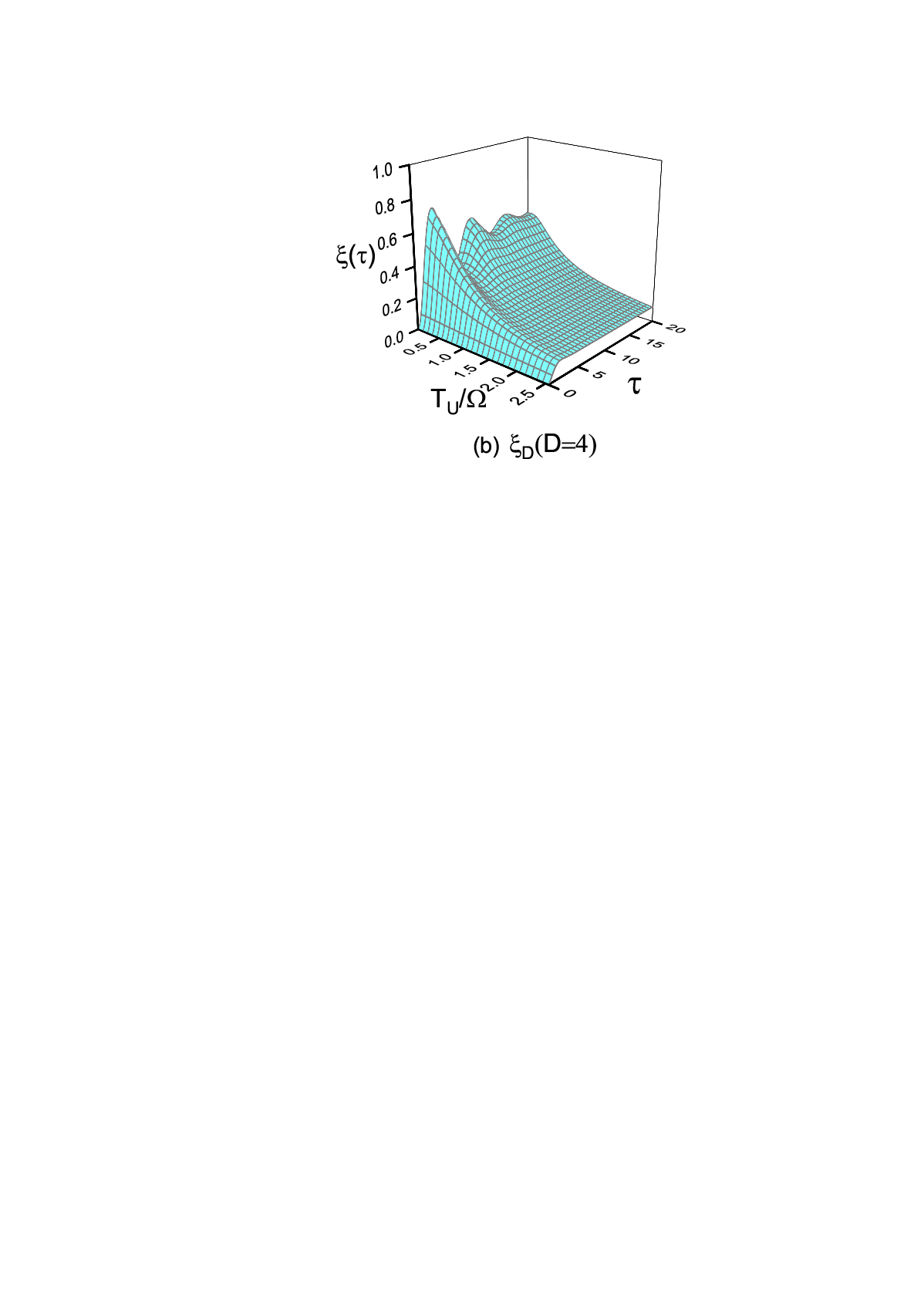}
\end{figure}

\begin{figure}
  \centering
  \includegraphics[width=1.0\textwidth]{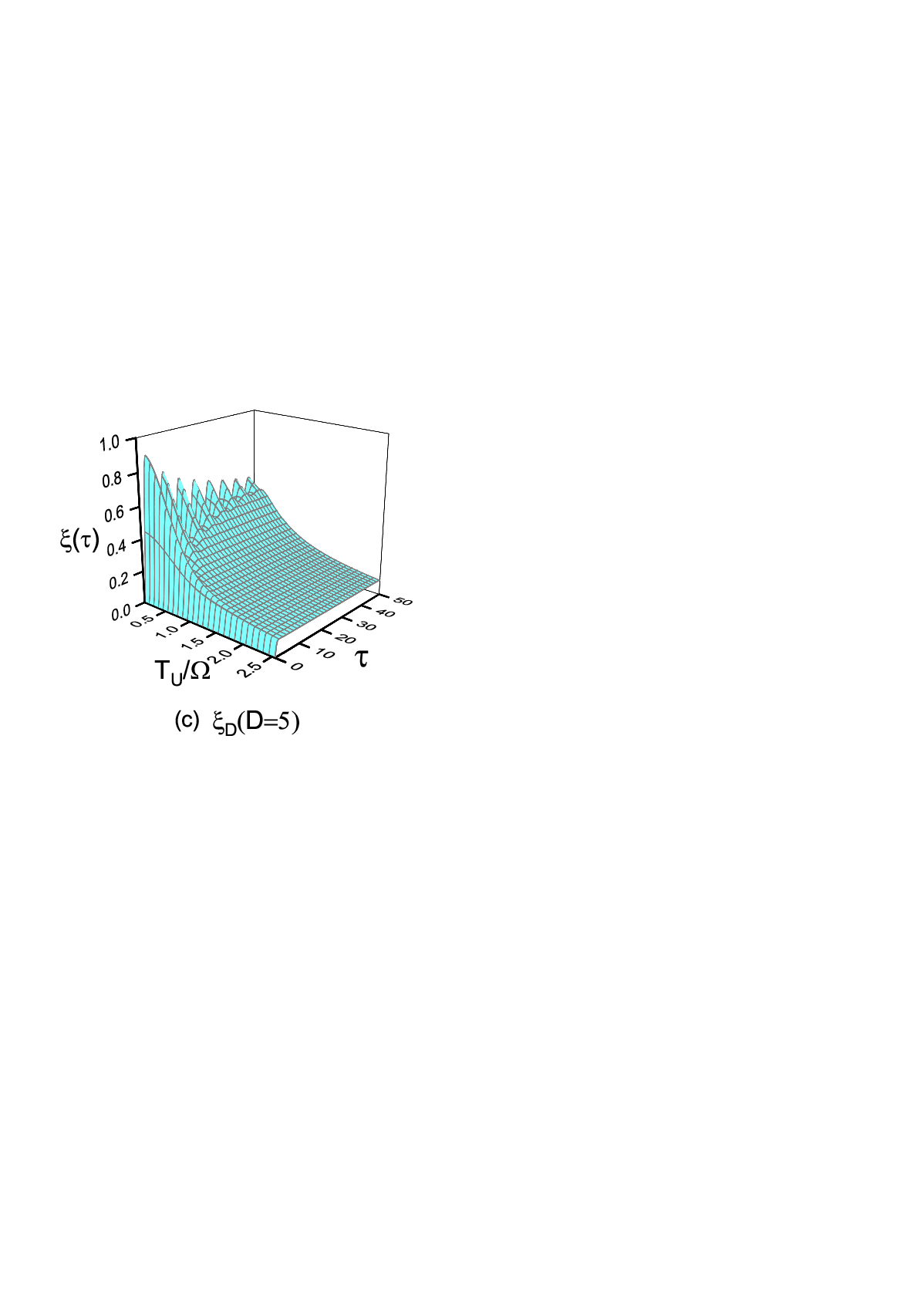}
\end{figure}

\begin{figure}
  \centering
  \includegraphics[width=1.0\textwidth]{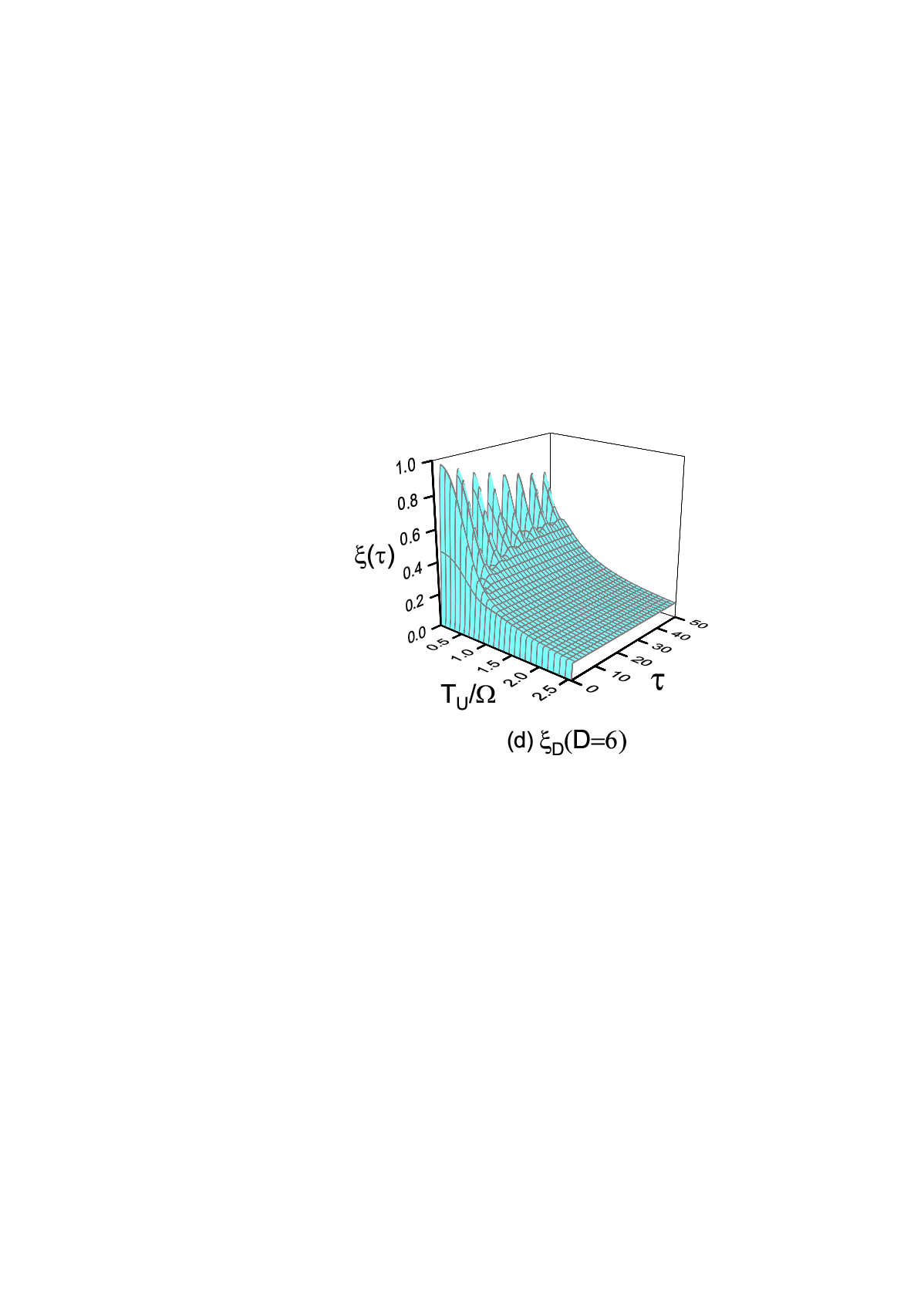}
\end{figure}

\begin{figure}
  \centering
  \includegraphics[width=1.0\textwidth]{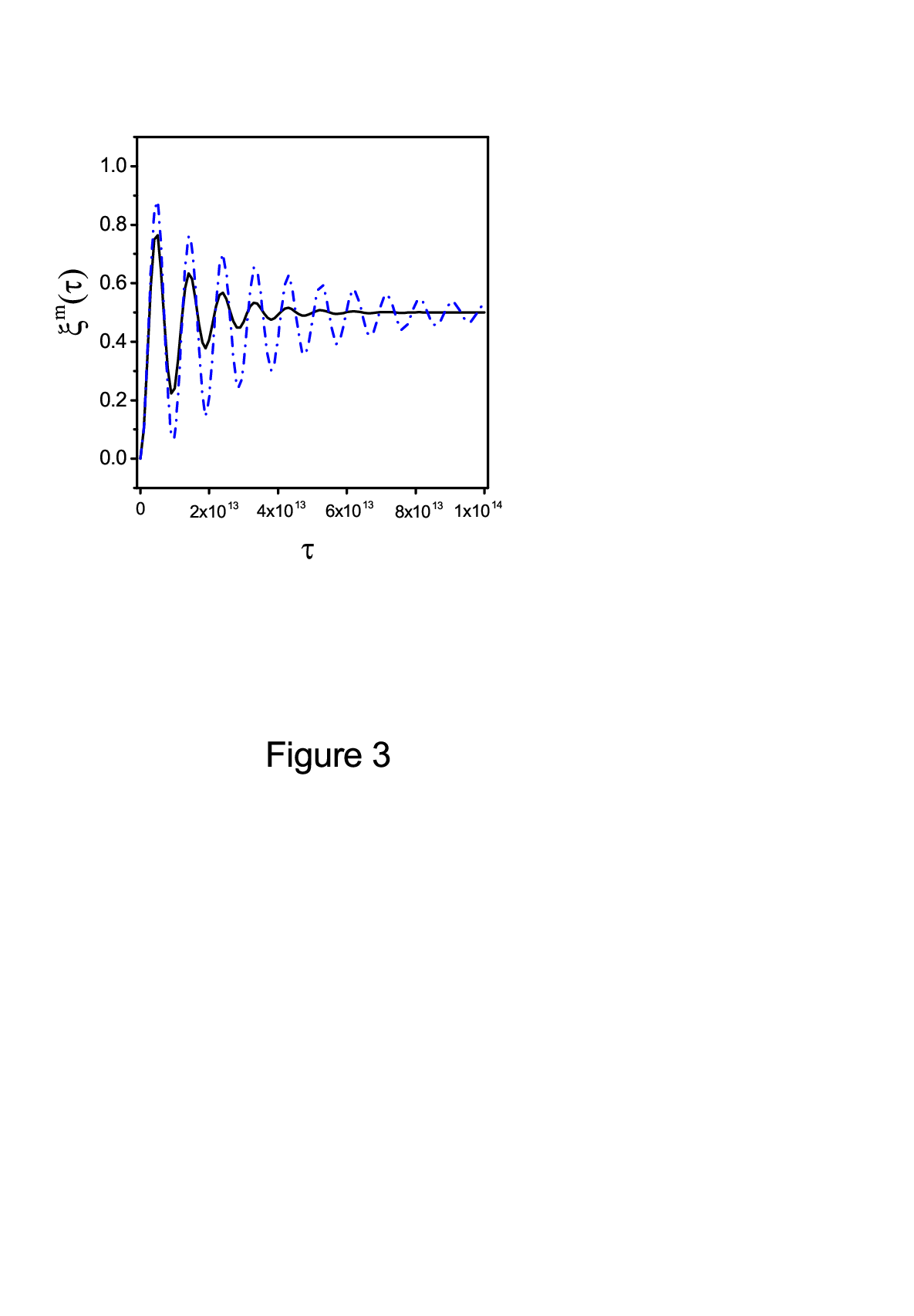}
\end{figure}

\end{document}